%%%%%%%%%%%%%%%%%%%%%%%%%%%%%%%%%%%%%%%%%%%%%%%%%%%%%%%
%%                (Plain) TeX file for               %%
%%                                                   %%
%%                                                   %%
%%          Infra-Red Finite Charge Propagation      %%
%%                                                   %%
%%                     by                            %%
%%                                                   %%
%%  E. Bagan, B. Fiol, M. Lavelle and D. McMullan    %%
%%                                                   %%
%%%%%%%%%%%%%%%%%%%%%%%%%%%%%%%%%%%%%%%%%%%%%%%%%%%%%%%
%%%%%%%%%%%%%%%%%%%%%%%%%%%%%%%%%%%%%%%%%%%%%%%%%%%%%%%

  %First some fonts

\font\bigbold=cmbx12
\font\eightrm=cmr8
\font\sixrm=cmr6
\font\fiverm=cmr5
\font\eightbf=cmbx8
\font\sixbf=cmbx6
\font\fivebf=cmbx5
\font\eighti=cmmi8  \skewchar\eighti='177
\font\sixi=cmmi6    \skewchar\sixi='177
\font\fivei=cmmi5
\font\eightsy=cmsy8 \skewchar\eightsy='60
\font\sixsy=cmsy6   \skewchar\sixsy='60
\font\fivesy=cmsy5
\font\eightit=cmti8
\font\eightsl=cmsl8
\font\eighttt=cmtt8
\font\tenfrak=eufm10
\font\sevenfrak=eufm7
\font\fivefrak=eufm5
\font\tenbb=msbm10
\font\sevenbb=msbm7
\font\fivebb=msbm5
\font\tensmc=cmcsc10
\font\tencmmib=cmmib10  \skewchar\tencmmib='177
\font\sevencmmib=cmmib10 at 7pt \skewchar\sevencmmib='177
\font\fivecmmib=cmmib10 at 5pt \skewchar\fivecmmib='177
%Some Families

\newfam\bbfam
\textfont\bbfam=\tenbb
\scriptfont\bbfam=\sevenbb
\scriptscriptfont\bbfam=\fivebb

\newfam\frakfam
\textfont\frakfam=\tenfrak
\scriptfont\frakfam=\sevenfrak
\scriptscriptfont\frakfam=\fivefrak

\newfam\cmmibfam
\textfont\cmmibfam=\tencmmib
\scriptfont\cmmibfam=\sevencmmib
\scriptscriptfont\cmmibfam=\fivecmmib
\def\bold#1{\fam\cmmibfam\relax#1}

%Definition of 8 point

\def\eightpoint{%
\textfont0=\eightrm   \scriptfont0=\sixrm
\scriptscriptfont0=\fiverm  \def\rm{\fam0\eightrm}%
\textfont1=\eighti   \scriptfont1=\sixi
\scriptscriptfont1=\fivei  \def\oldstyle{\fam1\eighti}%
\textfont2=\eightsy   \scriptfont2=\sixsy
\scriptscriptfont2=\fivesy
\textfont\itfam=\eightit  \def\it{\fam\itfam\eightit}%
\textfont\slfam=\eightsl  \def\sl{\fam\slfam\eightsl}%
\textfont\ttfam=\eighttt  \def\tt{\fam\ttfam\eighttt}%
\textfont\bffam=\eightbf   \scriptfont\bffam=\sixbf
\scriptscriptfont\bffam=\fivebf  \def\bf{\fam\bffam\eightbf}%
\abovedisplayskip=9pt plus 2pt minus 6pt
\belowdisplayskip=\abovedisplayskip
\abovedisplayshortskip=0pt plus 2pt
\belowdisplayshortskip=5pt plus2pt minus 3pt
\smallskipamount=2pt plus 1pt minus 1pt
\medskipamount=4pt plus 2pt minus 2pt
\bigskipamount=9pt plus4pt minus 4pt
\setbox\strutbox=\hbox{\vrule height 7pt depth 2pt width 0pt}%
\normalbaselineskip=9pt \normalbaselines
\rm}

%More general stuff

\def\pagewidth#1{\hsize= #1}
\def\pageheight#1{\vsize= #1}
\def\hcorrection#1{\advance\hoffset by #1}
\def\vcorrection#1{\advance\voffset by #1}

\newcount\notenumber  \notenumber=1              %Numbering does
\newif\iftitlepage   \titlepagetrue              %not start on title
\newtoks\titlepagefoot     \titlepagefoot={\hfil}%page
\newtoks\otherpagesfoot    \otherpagesfoot={\hfil\tenrm\folio\hfil}
\footline={\iftitlepage\the\titlepagefoot\global\titlepagefalse
           \else\the\otherpagesfoot\fi}

\def\abstract#1{{\parindent=30pt\narrower\noindent\eightpoint\openup
2pt #1\par}}
\def\smc{\tensmc}

%A nicer footnote

\def\note#1{\unskip\footnote{$^{\the\notenumber}$}
{\eightpoint\openup 1pt
#1}\global\advance\notenumber by 1}

\def\frac#1#2{{#1\over#2}}

\def\({\left(}
\def\){\right)}
\def\<{\langle}
\def\>{\rangle}
   %Partial derivatives
\def\2pd#1#2#3{\frac{\partial^2#1}{\partial#2\partial#3}}

\def\sqr#1#2{{\vcenter{\vbox{\hrule height.#2pt
        \hbox{\vrule width.#2pt height#1pt \kern#1pt
           \vrule width.#2pt}
        \hrule height.#2pt}}}}

\def\ni{\noindent}
\def\lqq{\lq\lq}
\def\rqq{\rq\rq}
\def\slash{\!\!\!\!/}

%%% Macro to generate the equation #'s automatically.
%%% To use start each new section (eg 3) with the commands
%%% \secno=3 \meqno=1 :this will start the equations with (3.1)
%%% Then in place of \eqno(3.1) type \eqn\descriptivename . To refer
%%% back to the equation simply type (\descritivename)
%%% For the appendix set \secno=0, \appno=1\meqno=1 etc
%%% If there are no sections, then set \secno=0

\global\newcount\secno \global\secno=0
\global\newcount\meqno \global\meqno=1
\global\newcount\appno \global\appno=0
\newwrite\eqmac
\def\romappno{\ifcase\appno\or A\or B\or C\or D\or E\or F\or G\or H
\or I\or J\or K\or L\or M\or N\or O\or P\or Q\or R\or S\or T\or U\or
V\or W\or X\or Y\or Z\fi}
\def\eqn#1{
        \ifnum\secno>0
            \eqno(\the\secno.\the\meqno)\xdef#1{\the\secno.\the\meqno}
          \else\ifnum\appno>0
            \eqno({\rm\romappno}.\the\meqno)\xdef#1{{\rm\romappno}.\the\meqno}
          \else
            \eqno(\the\meqno)\xdef#1{\the\meqno}
          \fi
        \fi
\global\advance\meqno by1 }

\def\eqnn#1{
        \ifnum\secno>0
            (\the\secno.\the\meqno)\xdef#1{\the\secno.\the\meqno}
          \else\ifnum\appno>0
            \eqno({\rm\romappno}.\the\meqno)\xdef#1{{\rm\romappno}.\the\meqno}
          \else
            (\the\meqno)\xdef#1{\the\meqno}
          \fi
        \fi
\global\advance\meqno by1 }
%%% Macro to assist in the references
%%% At the begining of the paper list the references in the order
%%% that they appear by the command \refn
%%% So if the first reference is to be
%%%  D. McMullan and I. Tsutsui Nucl. Phys. B121 (1994) 12
%%% then type \refn\us{D. McMullan and I. Tsutsui\np{121}{94}{12}}
%%% In the text this is simply refered to by [\us].
%%% At the end of the text type \listrefs

\global\newcount\refno
\global\refno=1 \newwrite\reffile
\newwrite\refmac
\newlinechar=`\^^J
\def\ref#1#2{\the\refno\nref#1{#2}}
\def\nref#1#2{\xdef#1{\the\refno}
\ifnum\refno=1\immediate\openout\reffile=refs.tmp\fi
\immediate\write\reffile{
     \noexpand\item{[\noexpand#1]\ }#2\noexpand\nobreak.}
     \immediate\write\refmac{\def\noexpand#1{\the\refno}}
   \global\advance\refno by1}
\def\semi{;\hfil\noexpand\break ^^J}
\def\nl{\hfil\noexpand\break ^^J}
\def\refn#1#2{\nref#1{#2}}

\def\up#1{$^{[#1]}$}

\def\jmp#1#2#3{{\it J. Math. Phys.} {\bf {#1}} (19{#2}) #3}
\def\ijmp#1#2#3{{\it Int. J. Mod. Phys.} {\bf A{#1}} (19{#2}) #3}

\def\pl#1#2#3{{\it Phys. Lett.} {\bf {#1}B} (19{#2}) #3}
\def\np#1#2#3{{\it Nucl. Phys.} {\bf B{#1}} (19{#2}) #3}

\def\pr#1#2#3{{\it Phys. Rev.} {\bf {#1}} (19{#2}) #3}

\def\prD#1#2#3{{\it Phys. Rev.} {\bf D{#1}} (19{#2}) #3}

\def\ann#1#2#3{{\it Ann. Phys.} {\bf {#1}} (19{#2}) #3}
\def\prp#1#2#3{{\it Phys. Rep.} {\bf {#1}C} (19{#2}) #3}

\def\zpC#1#2#3{{\it Z. Phys.} {\bf C{#1}} (19{#2}) #3}

\def\cjp#1#2#3{{\it Can. J. Phys.} {\bf{#1}} (19{#2}) #3}

%% Some Macros specific to this note%%

\def\a{\alpha}
\def\b{\beta}

\def\ga{\gamma}

\def\ep{\epsilon}
\def\eps{\epsilon}
\def\om{\omega}

\def\d{\delta}

\def\p{\pi^0}

\def\pa{\partial}

\def\vv{{\bold v}}
\def\al{\alpha}
\def\be{\beta}

\def\cov{(p-k)^2-m^2}

\def\de{\delta}
\def\De{\Delta}
\def\Dx{ {dx\over \sqrt{1-x}\sqrt{1-\vv{}^2 x}}}

\def\noncov{k^2-(k\cdot\eta)^2+(k\cdot v)^2}
\def\p{\partial}
\def\pn{p\cdot\eta}
\def\pv{p\cdot v}
\def\slash{\!\!\!/\,}

%%% Some parameters for this note %%%

\pageheight{24cm}
\pagewidth{15.5cm}
%\hcorrection{-2.5mm}
\magnification \magstep1
\voffset=8truemm
\baselineskip=16pt
\parskip=5pt plus 1pt minus 1pt

%%% The equations

\secno=0

%%% The references
\def\fop#1#2#3{{\it Found. of Phys.}  {\bf #1} {(19#2)} {#3}}
{\eightpoint
\refn\HATFIELD{This is reviewed in many elementary textbooks. See,
e.g., Sect. 17.3 of
B. Hatfield, {\sl Quantum Field Theory of Point Particles and
Strings}, (Addison-Wesley, Redwood City 1992)}
\refn\TOM{For a functional treatment of this, see
J. Breckinridge, M. Lavelle and T.G.
Steele, \zpC{65}{95}{155}}
\refn\NAKAOJIMA{N. Nakanishi and I. Ojima, {\sl Covariant Operator
Formalism of Gauge Theories and Quantum Gravity} (World Scientific,
Singapore 1990)}
\refn\YENNIE{H.M. Fried and D.R. Yennie, \pr{112}{58}{1391}}
\refn\MORCHIOSTROCCHI{G.\ Morchio and F.\ Strocchi, in {\sl Fundamental
Problems of Gauge Field Theory}, Ed.'s G.\ Velo and A.\ Wightman,
(Plenum Publishing Corporation 1986)}
\refn\COULFIRST{K. Johnson, \ann{10}{60}{536}}
\refn\HAGEN{R. Hagen, \pr{130}{63}{813}}
\refn\HECKA{D. Heckathorn, \np{156}{79}{328}}
\refn\ADKINS{G.S. Adkins, \prD{27}{83}{1814}}
\refn\MONSTER{M. Lavelle
and D. McMullan, \prp{279}{97}1}
\refn\QCDPOT{T.\ Appelquist, M.\ Dine and I.J.\ Muzinich,
\prD{17}{78}{2074}}
\refn\BLM{E. Bagan, M. Lavelle and D. McMullan, to appear in {\it 
Physical Review\/} {\bf D}}
\refn\WEINBERG{S.\ Weinberg, {\sl The Quantum Theory of Fields:
Vol. I\/}
(Cambridge University Press, Cambridge 1995), page 535}
\refn\DIRAC{P.A.M. Dirac, \lqq Principles of Quantum Mechanics\rqq,
(OUP, Oxford, 1958), page 302;
\cjp{33}{55}{650}}
\refn\HALL{K. Haller and E. Lim-Lombridas, \fop{24}{94}{217}}
\refn\NONLOC{F. Strocchi and A.S. Wightman, \jmp{15}{74}{2198}}
\refn\MAISON{D. Maison and D. Zwanziger, \np{91}{75}{425}}
\refn\NONLOCAL{J. Fr\"ohlich, G. Morchio and F. Strocchi,
\ann{119}{79}{241}}
\refn\BUCHH{D. Buchholz, \pl{174}{86}{331}}
\refn\BUCHH{L. Lusanna, \ijmp{10}{95}{3675}}
\refn\NONCOVAR{R. Haag, {\sl Local Quantum Physics}, (Springer-Verlag,
Berlin 1993)}
\refn\DOLLARD{J.D. Dollard, \jmp{5}{64}{729}}
\refn\JACKIW{R. Jackiw and L. Soloviev, \pr{173}{68}{1485}} 
}
%
%the beginning
%
%\leftline{\bf DRAFT  VERSION}
\rightline {UAB-FT-404}
\rightline {PLY-MS-96-03}
\vskip 38pt
\centerline{\bigbold INFRA-RED FINITE CHARGE PROPAGATION}
\vskip 28pt
\centerline{\smc Emili Bagan,{\hbox {$^1$}} Bartomeu Fiol,{\hbox
{$^{1}$}}\footnote{*}{{{{{\eightpoint{Current
address: Dept.\ of Physics and
Astronomy, Rutgers University, Piscataway, NJ 08855-0849}}}}}}
Martin Lavelle{\hbox {$^1$}} and  David McMullan{\hbox {$^2$}}}
\vskip 15pt
{\baselineskip 12pt \centerline{\null$^1$Grup de F\'\i sica Te\`orica
and IFAE}
\centerline{Edificio Cn}
\centerline{Universitat Aut\`onoma de Barcelona}
\centerline{E-08193 Bellaterra (Barcelona)}
\centerline{Spain}
\centerline{email: bagan@ifae.es}
\centerline{email: lavelle@ifae.es}
\vskip 13pt
\centerline{\null$^{2}$School of Mathematics and Statistics}
\centerline{University of Plymouth}
\centerline{Drake Circus, Plymouth, Devon PL4 8AA}
\centerline{U.K.}
\centerline{email: d.mcmullan@plymouth.ac.uk}}
%\vskip 7truemm
\vskip 35pt
{\baselineskip=13pt\parindent=0.58in\narrower\ni{\bf Abstract}\quad
The Coulomb gauge has a long history and many  uses.  It
is especially useful in bound state applications. An important
feature of this gauge is that the matter fields have an infra-red
finite propagator in an on-shell renormalisation scheme.
This is, however, only the case if the renormalisation point is chosen 
to be the static point on the 
mass-shell, $p=(m, 0, 0, 0)$. In
this letter we show how to extend this key property of the Coulomb
gauge to an arbitrary relativistic renormalisation point. This is
achieved through the introduction of a new class of gauges of which
the Coulomb gauge is a limiting case. A physical explanation for this
result is given.\par}
\bigskip\bigskip
\centerline{{\it To appear in Modern Physics Letters A}}

\vfill\eject
\noindent
In this letter we will study the  propagation
of a charged particle, such as an electron, in
a mass-shell renormalisation scheme.
To motivate this study, let us first recall\up{\HATFIELD}
the details of this calculation in a Lorentz gauge
with gauge parameter $\xi$.
Two renormalisation constants must
be introduced: a mass shift, $m\to m-\delta m$, and a fermion wave
function renormalisation,
$\psi\to\sqrt{Z_2}\psi$.
The mass shift, which we find by
requiring the presence of a pole at the physical mass, is found to
be gauge parameter independent as one would expect\up{\TOM} since the
electron mass is physical. Problems
arise when one now tries to demand that
the residue of the pole be unity: the $Z_2$ renormalisation constant
depends on the unphysical gauge parameter $\xi$ and
has in general an infra-red divergence, which obscures the physical
content of the theory.

Another way to see that there is a problem is to consider the
general form of the propagator around the mass shell.
Renormalisation group arguments\up{\NAKAOJIMA} indicate that the
one-loop renormalised propagator near the mass shell has the form
$$
\frac{(p^2-m^2)^{\beta}}{p_\mu\gamma^\mu-m}\,,\qquad {\rm
with}\quad \beta=\frac{e^2(-\xi-3)}{8\pi^2}
\,,
\eqn\poled
$$
and we see that
a pole structure emerges only in the Yennie gauge\up{\YENNIE}.
However, even for this gauge the resulting propagator cannot be identified
with that
describing charge propagation\up{\MORCHIOSTROCCHI}.

A gauge which appears not to have these problems is the Coulomb
gauge (for various treatments of QED in this gauge see Ref.'s
\COULFIRST-\ADKINS). Here one obtains the same mass shift as in
covariant gauges and $Z_2$ is infra-red finite. There is, however, an
often unappreciated subtlety here: infra-red finiteness only holds if
we are at the static point on the mass shell, i.e., if we demand
$p=(m,0,0,0)$. Details of this calculation can be found in Sect.\ 6 of
Ref.\ \MONSTER. The form of the propagator in Coulomb gauge near the
mass shell is quoted in Ref.~\MORCHIOSTROCCHI. Although this, like
(\poled), is in
general not a simple pole, in the {\it static limit\/} their formula
indeed reduces to
a pole.

Such kinematical configurations naturally
arise in many bound state problems. Indeed, even in QCD, the utility of
this
gauge in the calculation
of the static inter-quark potential is well known (see,
for example Ref.\ \QCDPOT).
A generalisation of the Coulomb gauge that allowed us to perform an
on-shell renormalisation of the electron propagator at an arbitrary,
relativistic point on the mass shell, i.e., for $p=m\gamma(1, {\bold
v})$  where $\gamma=1/\sqrt{1-{\bold v}^2}$,
would improve our understanding of these fundamental topics,
help extend the feasibility of such bound state calculations and
could be of use in the heavy quark effective theory where the heavy
quark has a well-defined velocity. We will provide such a
generalisation below.

\bigskip
Our class of gauges, adapted to motion in the $x^1$-direction,
is described by the condition
$$
\gamma^{-2}\p_1 A_1+\p_2 A_2+\p_3 A_3
+v^1[\p_0 A_1-\p_1 A_0] =0
\,,
\eqn\gauge
$$
where $v=(0,v^1,0,0)$. We will explain the physical arguments
that lead to this gauge choice in the conclusion.
It is clear that in the limit $v\to 0$ this condition reduces to the
Coulomb gauge.  However, we note that this class of gauges may
{\it not\/} be reached by a boost from the Coulomb gauge.
The vector boson propagator in this gauge is
$$
\eqalign{
D^v_{\mu\nu}=& {1\over k^2} \Big\{\Big. -g_{\mu\nu}
   +{(1-\xi)k^2-[k\cdot(\eta- v)]^2 \ga^{-2}\over [\noncov]^2}
k_\mu k_\nu \cr & \qquad\quad
-{k\cdot(\eta- v)\over\noncov
}\left[k_\mu (\eta+v)_\nu+(\eta+v)_\mu k_\nu\right]
\Big.\Big\}
\,,}
\eqn\propagator
$$
where $\xi$ is a (smearing) gauge parameter that we set to zero in what
follows and $\eta=(1,0,0,0)$ is a unit temporal vector.
We are not aware of work on this class of gauges previous
to our studies.

The electron propagator in this gauge may now be directly calculated.
The self-energy has the form
$$
\eqalign{
-i\Sigma =& e^2 \int\frac{d^{2\omega}k}{(2\pi)^{2\omega}} \Bigg\{\Bigg.
{1\over k^2}  {1\over \cov }
\Big[ 2(\om-1) p\slash - 2\om m -2(\om-1) k\slash \Big]  \cr
+& {1\over k^2}{1\over \cov}   \Big[-2(p\slash -m) \Big] \cr
+&{1\over \cov}{1\over \noncov} \Big[ 2(p\slash-m)+(\eta\slash+v\slash)
             \;  k\cdot (\eta-v)\Big] \cr
+&{1\over \cov}{1\over[ \noncov]^2}(p\slash-m) \Big[
\ga^{-2}(k\cdot\eta-k\cdot v)^2 - k^2 \Big] \cr
+& {1\over k^2}{1\over \cov}{1\over\noncov}\cr
&\quad \times
\Big[-(p^2-m^2)(\eta\slash+v\slash)\; k\cdot(\eta-v)-2k\slash\;
k\cdot(\eta-v)\; p\cdot (\eta+v)\Big] \cr
+& \Bigg. {1\over k^2}{1\over \cov}{1\over[\noncov]^2 }
(p^2-m^2) k\slash \Big[k^2-\ga^{-2}(k\cdot\eta-k\cdot v)^2\Big]
\Bigg\}\,.\!\!\!\!\!\!\!\!\!\!\!\!\!\!\!\!\!\!\!
}
\eqn\selfenergy
$$
This must be renormalised. With a little effort it may be seen that a
simple multiplicative renormalisation is inadequate to the task. In
this non-covariant gauge, this is not surprising. We find,
however, that a {\it matrix multiplication} renormalisation scheme is
appropriate. We use
$$
\psi\to\sqrt{Z_2}\exp\left\{-i\frac{Z'}{Z_2}\sigma_{\mu\nu}
\eta^\mu v^\nu\right\}\psi\,,
\eqn\matrixscheme
$$
which  is reminiscent
of a naive Lorentz boost upon the fermion. We find
it surprising and gratifying that this scheme is capable of
multiplicatively renormalising the propagator in this highly
non-covariant gauge. Note that we now have three renormalisation
constants: $\delta m$, $Z'$ and $Z_2$. The possible counterterms are
now (with $Z_2=1+\d Z_2$)
$$
-i\Sigma^{\rm counter}=i\d Z_2(p\slash-m)+ 2iZ'(p\cdot\eta v\slash
-p\cdot v\eta\slash)+i\d m\,.
\eqn\counterterms
$$
and the ultra-violet divergences of the self-energy are found to have a
similar form\note{The integration procedures and detailed results
will be presented elsewhere\up{\BLM}.}
$$
\eqalign{
-i\Sigma^{{\rm UV}}=i{\al\over4\pi}{1\over{2-\omega}}\Bigg\{\big.
-3m+&
(p\slash-m)\left[-3-2\chi({\bold v})\right]\cr
&\quad+2(\pv\; \eta\slash-\pn\; v\slash)\left[
{1\over\vv{}^2}+{1+\vv{}^2\over 2 \vv^2}\chi({\bold v})\right]
\big.\Bigg\}\,,
}
\eqn\uvdiv
$$
in $2\omega$ dimensions with $\a=(m^2)^{\omega-2}{e^2}/{4\pi}$
and  $\chi({\bold v})={\vert
{\bold v}\vert}^{-1}{\rm ln}\{\(1-\vert{\bold v}
\vert\)/\(1+\vert{\bold v}\vert\)\} $.

The renormalised self-energy (i.e., including both loops and counter
terms) may be written as
$$
-i\Sigma=m\a +p\slash\b +p\cdot\eta\slash \d +mv\slash\eps\,,
\eqn\renselfE
$$
where $\a,\dots, \epsilon$ are functions depending on $p^2$,
$p\cdot\eta$, $p\cdot v$ and $v^2$. Insisting that there is a pole at
the physical mass, $m$, fixes the mass shift condition. We find that
with $p^2=m^2$
for {\it any} values of $p\cdot\eta$, $p\cdot v$ and $v^2$ the
condition
$$
\tilde \a +\tilde \b +\frac{\( p\cdot\eta\)^2}{m^2}\tilde \d +
\frac{p\cdot v}{m}\tilde \eps=0\,,
\eqn\massshiftcond
$$
(the tildes signify that we put the momentum
$p^2$ on shell in the self-energy: $p^2=m^2$)
yields $ \d m={\a}( 3/{\hat \varepsilon} +4)m/(4\pi)$
which is just
the standard result for an arbitrary Lorentz gauge. (Note
that $1/\hat\epsilon = 1/(2-\omega) -\gamma_{\rm E}+{\rm ln}\,4\pi$.)
This gauge invariant result provides a check on our calculations.

In the on-shell scheme we now must demand that the residue of
this pole is unity. This we recall is where infra-red divergences may
lurk. It appears, however, as though non-covariance will immediately
lead to other problems.  This is because requiring that the
renormalised propagator has the form of the tree level propagator
now yields three equations
$$
\eqalign{
i\de Z_2-2\vv{}^2\;iZ'&=\bar\de_L-\bar\be_L-2m^2\bar\Delta\,,\cr
i\de Z_2-
\phantom{\vv{}^2} 2\;iZ'&=\ga^{-1} \bar\ep_L-\bar\be_L-
2m^2\bar\Delta\,,\cr
i\de Z_2\;\,\phantom{-2\vv{}^2\;iZ'}&=-\bar\al_L-2\bar\be_L-2m^2
\bar\Delta\,,
}
\eqn\threeeqs
$$
where
$$
\Delta(p\cdot\eta, p\cdot v,v^2) =
\Big(\frac{\pa\a}{\pa p^2} + \frac{\pa\b}{\pa p^2}
+\frac{(p\cdot\eta)^2}{m^2}\frac{\pa\d}{\pa p^2} +\frac{p\cdot v}{m}
\frac{\pa\eps}{\pa p^2}\Big)
\,,
\eqn\XZX
$$
and the bars denote that the functions are evaluated
at the point on the
mass shell given by, $p=m\gamma(\eta+v)$. Note further that the $L$
subscript signifies that the functions $\a_L$ etc.\
only contain loop and mass shift contributions.
Since these are three equations and we now have only two unknowns, $Z'$
and $Z_2$, we
cannot in general proceed any further. However, at our chosen
point on the
mass shell we find that these equations
reduce to two independent ones and our counterterms are now uniquely
determined in terms of the loop contributions.

The result for $Z'$ can be seen to be free of infra-red divergences,
but that for $Z_2$ contains various infra-red divergent integrals.
However, we find that the
overall combination of such divergences is:
$$
\eqalign{
m^2\bar\De_{IR}=i{\al\over4\pi} \int_0^1 du\; u^{2\om-5}\Big\{-2&
\Big.
+2\int_0^1 \Dx \left[1+\vv{}^2-2\vv{}^2 x\right] \cr
-&
\Big.(1-\vv{}^2)\int_0^1 \Dx\; x{3+\vv{}^2-2\vv{}^2 x\over 2
(1-\vv{}^2 x)}
\Big\}
}
\,,
\eqn\no
$$
where infra-red finite terms are ignored. The integrals over $x$ now
yield exactly $+2$ and so the {\it infra-red divergences all cancel}.
We have thus performed a (matrix) multiplicative, infra-red finite,
on-shell renormalisation of the electron propagator in this class of
gauges. This concludes the first part of this letter.
\bigskip
We now want to consider scalar QED in these gauges.
It is known\up{\WEINBERG} that in covariant gauges the infrared structure
of the
propagator is independent of the spin of the field. Thus  the scalar
electron
propagator is also infra-red finite in the covariant Yennie gauge. Here,
in a non-covariant formalism, things are not so clear. Furthermore for a
scalar \lq electron\rq\ we cannot have a multiplicative matrix
renormalisation scheme such as Eq.\ (\matrixscheme): this further
tightens the conditions our gauge must fulfill.

The procedure we have to follow should be clear: we use the same photon
propagator but a very different (scalar) electron propagator and
calculate the one-loop self-energy. (Note that the tadpole diagram
vanishes.) We demand a pole at the physical mass and that its residue
is unity. Our wave function renormalisation is now just $\phi\to
\sqrt{Z_2}\phi$.

The mass shift has again a gauge invariant value in this theory. The
condition that the pole has unit residue now implies just one equation
since we do not have any gamma matrix structure and we find
$$
\delta Z_2=\frac\alpha{4\pi}\left\{ (6+2\chi(v))\frac1{\hat\epsilon}
+4\left(1-\gamma^{-2}\chi(v)-
\frac1{|{\bold v}|}\left[
L_2(|{\bold v}|)-L_2(-|{\bold v}|)
\right]
\right)\right\}\,,
\eqn\ztwoscal
$$
with $L_2$ being the dilogarithm.
Once again {\it all the infra-red divergences have cancelled} and  an
on-shell, multiplicative renormalisation has been achieved.
This is an important check of our earlier calculation and shows the
power of the gauge we have introduced.  We finally note that
any attempt to renormalise at a different point on the mass shell
leads to the appearance of infra-red divergences. This indicates the
sensitivity of the above cancellation of the infra-red singularities.

\bigskip

To conclude this letter let us offer an explanation of why these gauges
(parameterized by $v$) possess such attractive properties.
We have seen that in the gauge (\gauge) the propagator for the
matter field (either scalar or spinor) is infra-red
finite at the appropriate (moving) point on the 
mass shell. We recall that the
infra-red problem reflects the fact that physical
charged particles are always dressed with an electromagnetic cloud
which the Lagrangian fermion or scalar field
does not properly reflect --- unless, as we now argue,
our gauge is used.

We remember that we
have also noted that the Yennie gauge propagator shares this
property for the whole of the mass shell. However, in that
case a connection with the propagation of charged particles
is not visible. For our class of gauges, though, such a
connection can be made\up{\MONSTER}.

Recall that
the gauge transformation that
takes an electron into Coulomb gauge is just
$$
%sign of e corrected here to agree with bigger paper!
\psi(x)\to\exp\(-ie\pa_iA_i/\nabla^2\)\psi(x)\,.
\eqn\coultrans
$$
The canonical commutator of the electric field with this transformed
object gives the Coulomb electric field expected of a static
charge\up{\DIRAC,\HALL}. The general form of this transformation is in accord
with what is known from rigorous studies\up{\NONLOC-\NONCOVAR,
\MONSTER} of
charged particles in gauge theories; i.e., that a charge is accompanied
by a non-local electromagnetic cloud whose non-covariant form reflects
the known difficulties in reconciling Lorentz transformations and gauge
transformations for non-local, non-observable quantities.
What we have used in this letter is a generalisation of Eq.~\coultrans,
i.e., the transformation into our gauge condition (\gauge) is such
that\up{\MONSTER} the canonical commutators
of the electric or magnetic fields with the
gauge transformed charged matter fields
give the results expected of a charge moving with velocity $v$. In
other words the charged fields in the gauge we have introduced are
automatically accompanied by the dressing which surrounds a
charged particle moving with velocity, $(v_1,0,0)$. The reason that the
gauge \lq works\rq\ for both fermions and scalars is that the magnetic
fields associated with the fermion's anomalous magnetic moment fall off
more rapidly than $1/r$ and thus these interactions do not lead to
infra-red problems\up{\DOLLARD}.

Although the argument used to construct the gauge choice (\gauge) is
essentially classical, we have seen that our quantum
calculations solidly back up the idea that such fields may play
the role of good asymptotic fields in gauge theory as was predicted in
Ref.\ \MONSTER. 

\bigskip
\ni {\bf 
Note added:} After this paper was completed we received a letter 
from R.\ Jackiw drawing our attention to his paper with L.\ 
Soloviev\up{\JACKIW}. Using 
the exponentiation of the soft divergences property found there, 
one can convince oneself that the infra-red finiteness property 
demonstrated here will hold at all orders. We thank Prof.\ Jackiw for 
this reference. The next stage in this programme is to consider 
vertices: here we must associate a different (velocity dependent) 
dressing to each charged leg and no gauge choice can remove all the 
dressings. Since we wrote this letter we have seen in 
explicit one-loop calculations that the soft divergences in the 
vertex can be removed by incorporating dressings. This and all-orders 
arguments will be 
presented elsewhere.  
\bigskip
\ni {\eightpoint 
{\bf Acknowledgements:} EB and BF received support from CICYT
research project AEN95-0815. BF also thanks the Generalitat de
Catalunya for a grant.
MJL thanks project CICYT-AEN95-0882 for
support and the University of Plymouth for their kind hospitality. DM
thanks George Leibbrandt for useful discussions on the
Coulomb gauge. MJL and DM thanks the organisers of the 1996
UK Theory Institute for providing them with a stimulating forum for
discussions. We thank R.\ Tarrach for stressing the importance of the
scalar QED calculation.}
\bigskip
\bigskip
\bigskip

  \vfill\eject\immediate\closeout\reffile%\parindent=20pt
  \centerline{{\bf References}}\bigskip\eightpoint\frenchspacing%
  \input refs.tmp\vfill\eject\nonfrenchspacing

\bye